# Choroidal thinning assessment through facial video analysis

Qinghua He[1], Yi Zhang[1], Mengxi Shen[2], Giovanni Gregori[2], Philip J. Rosenfeld[2] and Ruikang K. Wang[1, 3, *]

[1]Department of Bioengineering, University of Washington, Seattle, Washington 98105, USA
[2]Department of Ophthalmology, Bascom Palmer Eye Institute, University of Miami Miller School of Medicine, Miami, FL, USA
[3]Department of Ophthalmology, University of Washington, Seattle, Washington 98109, USA

**ABSTRACT**

Different features of skin are associated with various medical conditions and provide opportunities to evaluate and monitor body health. This study created a strategy to assess choroidal thinning through the video analysis of facial skin. Videos capturing the entire facial skin were collected from 48 participants with age-related macular degeneration (AMD) and 12 healthy individuals. These facial videos were analyzed using video-based trans-angiosomes imaging photoplethysmography (TaiPPG) to generate facial imaging biomarkers that were correlated with choroidal thickness (CT) measurements. The CT of all patients was determined using swept-source optical coherence tomography (SS-OCT). The results revealed the relationship between relative blood pulsation amplitude (BPA) in three typical facial angiosomes (cheek, side-forehead and mid-forehead) and the average macular CT ($r = 0.48$, $p < 0.001$; $r = -0.56$, $p < 0.001$; $r = -0.40$, $p < 0.01$). When considering a diagnostic threshold of 200μm, the newly developed facial video analysis tool effectively distinguished between cases of choroidal thinning and normal cases, yielding areas under the curve of 0.75, 0.79 and 0.69. These findings shed light on the connection between choroidal blood flow and facial skin hemodynamics, which suggests the potential for predicting vascular diseases through widely accessible skin imaging data.

**INTRODUCTION**

Situated between the retina and the sclera, the choroidal perfusion provides essential nutrients to the retinal pigment epithelium and the outer retina[1] and allows for the dissipation of heat from light impinging on the fundus[2]. Alterations in choroidal circulation have been associated with ocular health and visual function[3,4]. Studies investigating the relationship between choroidal volume and choroidal circulation suggest that choroidal thinning may be a marker of a compromised choroidal circulation[5]. Consequently, the evaluation of choroidal thinning has emerged as a critical area of focus,[6-8] and the assessment of choroidal thinning has become are area of interest in the diagnosis and management of conditions like age-related macular degeneration (AMD)[9], myopia[10], and diabetic retinopathy[11]. In such cases, changes in choroidal thickness (CT) have been correlated with the progression of the disease and are thought to be prognostic indicators.

Conventional assessments of CT have relied heavily on ophthalmic imaging techniques, with optical coherence tomography (OCT) being the most valuable [12,13]. While these techniques have proven invaluable within clinical context, their implementation requires specialized equipment, trained personnel, and time-intensive procedures. Moreover, their utility remains largely confined to controlled clinical environments, impeding their broader adoption for routine screening. Given the constraints imposed by existing assessment techniques, there is a need to create innovative tools that facilitate accessible, efficient, and widespread screening for choroidal thinning.

According to the angiosomes theory, specific anatomical regions within human tissues are supplied with blood through dedicated source arteries and accompanying veins[14,15]. The skin, being the body's largest organ, comprises a multitude of these angiosomes, intricately linked with the vascular network of the circulatory system[16,17]. This forms the basis for evaluating vascular conditions by examining hemodynamics within the skin. Previous research has hinted at the potential of harnessing camera data streams from skin to predict vascular conditions[18,19]. Integrating these advancements with the investigation of choroidal thinning may open the door to a transformative paradigm in the field.

This study uses facial video analysis to craft a tool for assessing choroidal thinning. Based on facial videos, we introduce the trans-angiosome imaging photoplethysmography (TaiPPG) strategy; a method that computes regional facial blood pulsation to create regional markers capable of predicting CT (Fig.1). The findings provide compelling evidence that relative blood pulsation amplitude (BPA) derived from facial skin angiosomes can predict the CT with notable accuracy. This validation attests to the efficacy of our analysis strategy and underscores its potential to enhance our ability to conveneiently assess choroidal thinning. This study describes a method that could equip healthcare professionals with advanced tools that bypass traditional imaging methods to screen for patients at risk for disease progression.

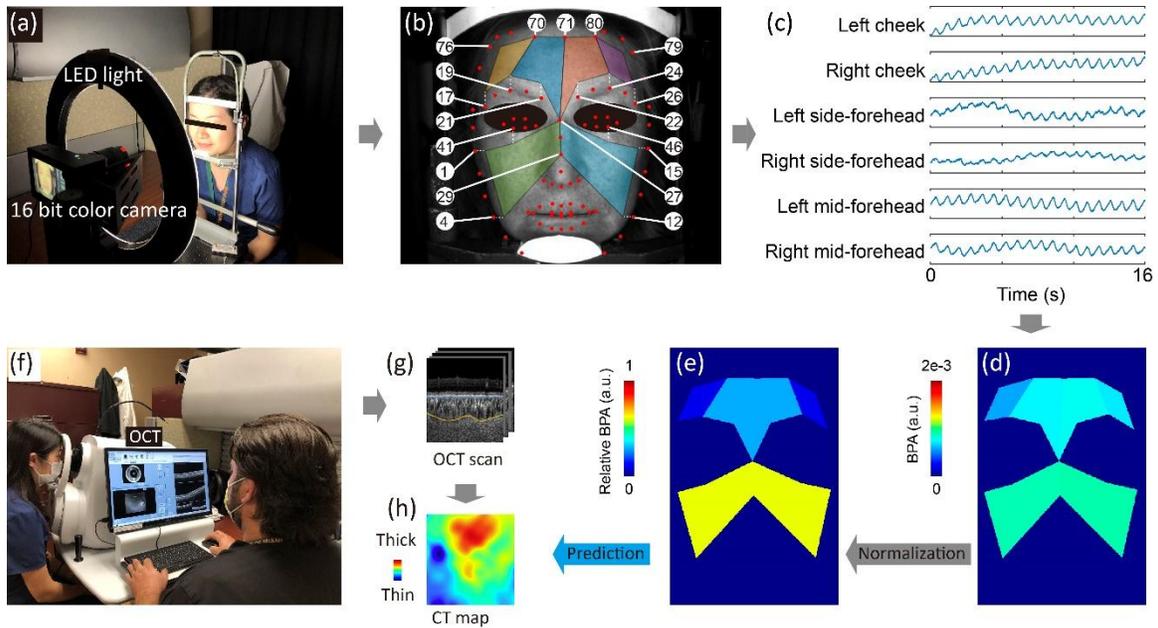

Fig. 1 The photograph (a) and workflow of trans-angiosome imaging photoplethysmography (TaiPPG) measurements in clinics. The facial skin was illuminated by a ring light emitting diode (LED) and imaged by a 16-bit camera. The regions of the cheek, side-forehead and mid-forehead were selected by the modified 81-points face landmark detection algorithm (b). The regional PPG pulses (c) were extracted from the video and processed with a lock-in amplification algorithm to calculate blood pulsation amplitude (BPA) (d) and normalized BPA (e). The photograph (f) of optical coherence tomography (OCT) scanning in clinics and workflow of using OCT B-scans (g) assembled into an *en face* projection map to measure the choroidal thickness (CT) (h). The details of workflow are described in the Methods section.

## RESULTS

### Participants

A total of 60 participants were enrolled, comprising 36 (60%) women. The recruitment of patients was primarily focused on populations with AMD, owing to the proposed importance of reduced choroidal circulation in disease progression[3,20]. Consequently, the study encompassed 33 subjects with dry AMD, 15 with wet AMD, and 12 exhibiting normal eyes (as presented in Table 1). Each participant underwent both facial video recording and swept-source OCT (SS-OCT) imaging. Multiple measurements were gathered from each AMD participant during successive follow-up visits. Normal cases underwent one measurement.

Table 1. Study population

| Patient number | Gender | | Age range (years) | | |
|---|---|---|---|---|---|
| | Female | Male | <60 | 60-79 | 80-89 |
| 60 | 36(60%) | 24(40%) | 22(37%) | 24(40%) | 14(23%) |
| Diagnosis | Dry AMD (n=33) | | Wet AMD (n=15) | | Normal (n=12) |
| Gender(Female/Male) | 23/10 | | 9/6 | | 6/6 |
| Age(years) | 76±10 | | 77±6 | | 67±10 |
| SBP(mmHg) | 133±14 | | 132±16 | | 135±18 |
| DBP(mmHg) | 72±7 | | 70±8 | | 75±8 |

*SBP/DBP systolic/diastolic blood pressure*

### Facial video indicators

To pride a visual depiction of the correlation between facial BPA and CT, we have chosen six illustrative cases and presented their outcomes in the following section. The visual representation in Fig. 2(a) highlights facial BPA maps processed from the pixelated pulse waveform and distinct relative BPA values across the cheek, side-forehead, and mid-forehead regions for these six individuals. A noticeable trend emerged from the leftmost to the rightmost cases, indicating a gradual decrease in BPA values in the forehead region,. Fig. 2(b) provides an insightful view of CT, revealing a discernible escalation in thickness as we progress from the left to the right across the subjects. This visual presentation underscores the potential relationship between facial BPA and CT, emphasizing the importance of considering regional variations in BPA values and their corresponding impact on CT. The selected cases serve as representative instances, shedding light on the nuanced patterns within facial BPA and CT, contributing to a deeper understanding of their interconnected dynamics.

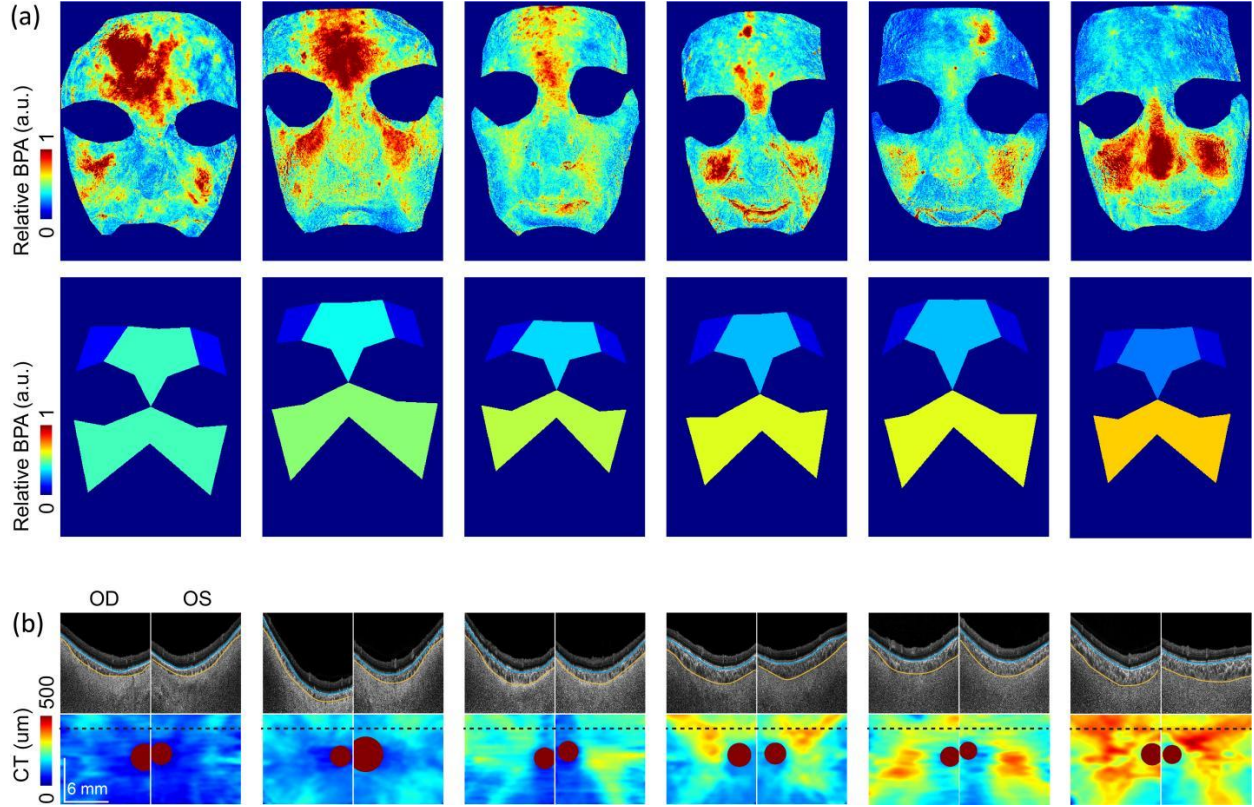

Fig.2 (a) Relative blood perfusion amplitude (BPA) map of six typical cases with relative cheek BPA values changing from low to high. Top row: pixelate related facial BPA maps; bottom row: relative BPA values in outlined central forehead, side forehead and cheek regions. (b) SS-OCT B-frames and CT maps of corresponding cases in (a). The choroid thickness (CT) maps are produced by measuring the distance between the Bruch's membrane (blue line) and choroid-sclera interface (yellow line). The representative B-frames were extracted from the positions labeled with black dotted lines in the CT maps. OD: Oculus Dexter; OS: Oculus Sinister.

**Relationship between video indicators and clinical measurements**

We found significant correlations between the mean CT measured from SS-OCT scanning and automatic video-based facial indicators describing cheek BPA ($r = 0.48$, $p < 0.001$), side forehead BPA ($r = -0.56$, $p<0.001$), and central forehead BPA ($r = -0.40$, $p<0.01$) (Fig. 4(a-c)). It is worth noting that the outliers were detected and masked in the correlation analysis. This method uses regression analysis to identify outliers based on standardized residuals. After importing the data and creating a scatter plot, a linear fit is applied. The standardized residual is calculated and checked to spot any outliers (values beyond ±2). Identified outliers are masked using a tool, which automatically updates the regression analysis, excluding these points for more accurate results. It is worth noting that the interrelation among the three facial indicators is attributable to the way the BPA normalization denominator is calculated, which involves summing the BPA values from the cheek, side forehead, and central forehead areas. Consequently, the observed correlations between CT and the facial indicators could be the result of interactions between the choroid and one or two specific facial regions, rather than all three..

Based on the scatter plots, three linear regression models were developed to predict the CT. Although there is no established standard to define choroidal thinning in clinics, we used an artificial threshold to assess the performance of our models in classification. We separated the 60 cases into choroidal thinning and non-choroidal thinning groups using a cutoff at 200 μm as the reference. We then used the relative BPA values at this cutoff as the prediction threshold to classify these cases again. The resulting classification accuracy for the three models is 69.5% (relative cheek BPA, sensitivity: 88.9%, specificity: 39.1%), and 69.5% (relative side forehead BPA, sensitivity: 86.1%, specificity: 45.4%), and 66.9% (relative side forehead BPA, sensitivity: 91.7%, specificity: 26.1%), respectively. Receiver operating characteristic (ROC) curves for these three models are presented in Fig. 4(d-f). The area under the ROC (AUC) values were calculated to be 0.75 ± 0.06, 0.79 ± 0.06 and 0.69 ± 0.07. These results suggest that video-based TaiPPG is capable of generating effective classification models for choroidal thinning.

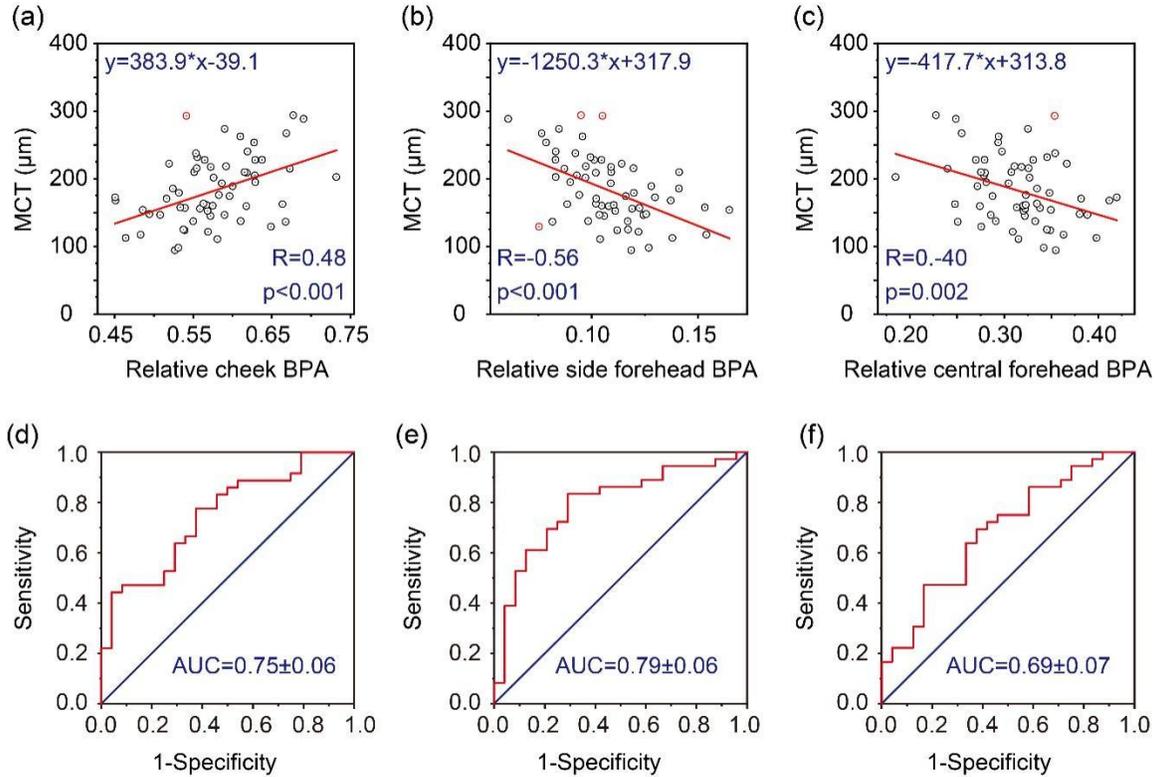

Fig.3 (a-c) Correlation plots between mean choroid thickness (MCT) and relative blood perfusion amplitude (BPA) values of cheek (a), side forehead (b) and central forehead (c). Outliers identified and masked due to standardized residuals exceeding 2 in the initial regression round.. (d-f) ROC curves between choroidal thinning and non-choroidal thinning groups using a cutoff at 200 μm as the reference, which were based on facial indicators from cheek (d), side forehead (e) and central forehead (f). ROC: receiver operating characteristic; AUC: area under the ROC curve.

**METHODS**

**Study design and participants**

The TaiPPG imaging method underwent clinical validation in a study that aimed to predict CT and detect choroidal thinning in patients with AMD. The study strictly adhered to the tenets of the Declaration of Helsinki and was carried out in accordance with the Health Insurance Portability and Accountability Act. Approval was obtained from the Institutional Review Board of the University of Miami. Facial videos were collected from the Bascom Palmer Eye Institute, University of Miami Miller School of Medicine, between 30 April 2021 and 10 August 2023. These videos were subsequently anonymized and de-identified to protect the privacy of the participants. The study included dry AMD, wet AMD, and normal cases, with exclusion criteria being uncorrectable global and local motion in the face, incomplete facial skin video coverage due to masks, hair, etc., overexposure of facial skin, and poor quality of OCT scanning. Throughout the data collection process, clinicians were masked to the technology used in video processing. The OCT and TaiPPG data were processed separately by different analysts, who were also masked to the data outside of their area of analysis.

**Extraction of facial indicators**

In the TaiPPG system, a 16-bit camera (Chronos 1.4; Kronos, Canada) was utilized to capture video images of reflections on the facial skin, which were produced by the illumination of a white ring LED light source. A zoom lens (Computar 12.5-75mm/f 1.2; CBC Group, Japan) was used to relay the facial image onto the camera sensor. To reduce specular reflections on the skin surface, we placed a pair of polarized films in front of the camera lens and light source at orthogonal angles. The U1/U2 equations of illumination uniformity were measured to be 0.99 and 0.95, respectively, indicating that the illumination was uniform across the facial skin. To prevent large motions, a chin rest was installed in front of the camera and light source to stabilize the patient's head.

Forty-eight patients diagnosed with AMD (Dry: 33; Wet: 15) and 12 normal cases were included in this study, and TaiPPG and OCTA imaging were used to follow them. Before recording the facial skin video in TaiPPG imaging, the subjects were asked to sit calmly in a chair for 30 seconds for heart rate stabilization. The room temperature and humidity were kept constant during the experiment. The video recording lasted for 16 seconds at 100 frames per second, with each frame having a resolution of 640 by 512 pixels in the R, G, and B channels. The acquired video was saved as 16-bit modes for further processing. To minimize motion artifacts, we applied a sub-pixel registration algorithm to co-register frames before extracting the pulsation signal[21,22]. The green channel was selected for analysis since it had the highest SNR among the three channels.

We employed a modified 81-point landmark detection algorithm to automatically identify facial landmarks[23]. This is a custom shape predictor model trained to find 81 facial feature landmarks that have been widely used in facial recognition applications[24-26]. Utilizing these landmarks, we delineated the cheek, side forehead, and central forehead regions, followed by the extraction of pulse waveform. Each pulse waveform underwent a lock-in amplification algorithm to derive the BPA[27]. Briefly, a global pulse wave will be extracted from the video cube. This will undergo heart rate filtering to form a reference function. The function will then be applied to other pulse waveforms to amplify the signal at the heart rate frequency and calculate their amplitudes. The resultant BPA map encapsulated the averaged magnitudes within the six outlined regions. To improve data presentation, we merged information from the left and right facial sides and normalized each region using overall face BPA values derived from the cheek and forehead regions. This process yielded three distinct groups of facial indicators: relative cheek BPA, relative side forehead BPA, and relative central forehead BPA.

**CT measurements**

The OCT imaging was performed using a SS-OCT system (PLEX Elite 9000; Carl Zeiss Meditec, CA) with a laser source centered at 1050 nm and a bandwidth of 100 nm. The system has an axial resolution of ~5 micrometers and a lateral resolution of approximately 20 μm in tissue. A 12 mm by 12 mm SS-OCT angiography (SS-OCTA) scan pattern centered on the fovea, consisting of 500 A-scans per B-scan and 500 B-scan positions with each B-scan position repeated twice, was used to image all subjects. From the three-dimensional scan volume, a choroidal data slabs were semi-automatically generated by segmenting the choroidal boundaries, including the Bruch's membrane and choroid-sclera interface, using a previously reported algorithm[28,29]. The *en face* CT map was then generated by measuring the distance between choroidal boundaries. The mean CT (MCT) value was obtained by averaging the CT map of both eyes (OS and OD).

**Statistical analysis**

Summary statistics are provided in the form of counts (percentages) and median (interquartile range) for categorical and continuous variables. To assess the relationships between CT and facial indicators (cheek, side forehead, and central forehead relative BPA), we utilized Pearson correlations. Subsequently, based on scatter plots, we constructed three linear regression models for predicting CT from facial indicators. Applying a threshold of 200 μm, we categorized 60 cases into choroidal thinning and non-choroidal thinning groups, enabling the calculation of prediction accuracy for the three models. Furthermore, we employed the area under the receiver operating characteristics (AUROC) to gauge model discrimination, presenting the AUROC along with its 95% confidence interval. All analyses were executed using Origin 2021b, and the significance level was set at an alpha of 0.05.

**DISCUSSION**

In this study, we introduced a method that utilized data streams acquired by a camera for the automated and objective evaluation of CT. Our approach involves the application of TaiPPG technology to detect relative BPA in specific facial regions, namely the cheek, side-forehead, and central forehead. Notably, this methodology achieved a commendable 65-70% accuracy in distinguishing choroidal thinning, positioning it as a promising screening strategy for recommending additional imaging technologies like OCT. What sets this approach apart is its potential application not only in the context of the assessment of choroidal thinning but also in the context of various ocular diseases. The utilization of video-based assessment not only enhances diagnostic capabilities but also offers the advantage of reducing the economic and physical burden on patients and healthcare systems during health monitoring. One of the key strengths lies in the ease of obtaining video recordings, a process that can be seamlessly executed by an operator with imaging expertise. This characteristic opens up the possibility of scaling up the method to cater to a larger population. The practicality of video-based assessment makes it particularly well-suited for integration into mobile and digital health applications, positioning it as a convenient and feasible diagnostic strategy for widespread use.

While the vascular explanation for this hemodynamic connection between the choroid and facial skin is not yet fully elucidated, we hypothesize that collateral circulation from the external carotid arteries (ECAs) might underlie this connection, similar to observations in cerebral studies where internal carotid stenosis induces increased blood supply in external carotid arteries to compensate for cerebral tissue ischemia[30,31]. As shown in Fig. 4, within the context of our study, we suggest that there may be an abnormal chronic reduction of blood supply in the arteries downstream along the OA, perhaps due to stenoses in the OA or carotid arteries, triggering a supplementary response that supports ocular tissue through the collateral circulation from the ECAs[32]. This is specifically mediated by the anastomoses between the downstream OA branches and their ECA branches, such as the dorsal nasal arteries (DNA) connecting to the angular termination of the facial artery (FA) and infraorbital arteries[33]. Consequently, as a result of decreased terminal OA perfusion to the forehead or cheek, there may be increased blood flow in the FA leading to heightened cheek blood pulsation, which would establish retrograde blood flow into the OA and a positive relationship between CT and relative cheek BPA. Conversely, if retrograde blood flow couldn't be established in the OA, then the BPA of the side and central forehead exhibit a negative correlation with CT suggesting that the decreased blood flow from the OA to the forehead results in compensatory bloodflow from the ECA. Specifically, a pronounced negative correlation between the side-forehead BPA and CT may be due to a relative increase to the forehead from the superficial temporal artery (STA) branches to compensate for the reduced perfusion from the OA to this same BPA, though this requires further investigation and substantiation.

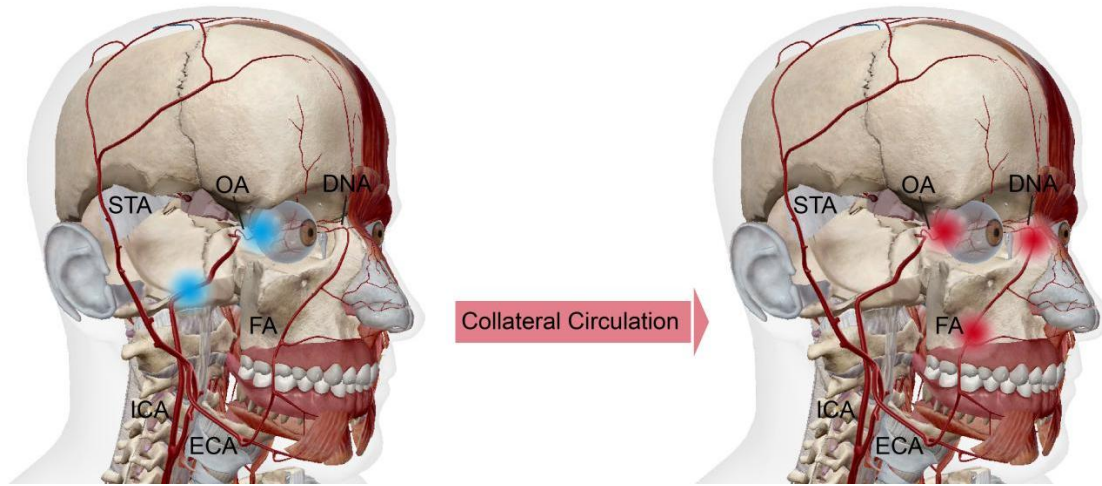

Fig.4 Illustration of our proposed collateral circulation mechanism depicting the correlation between CT and relative cheek BPA levels. Blue indicates decreased perfusion; Red indicates increased perfusion. OA: ophthalmic artery; FA: facial artery; STA: superficial temporal artery; DNA: dorsal nasal arteries; ICA:internal carotid artery; ECA: external carotid artery.

Our study introduces TaiPPG as a potential supplementary modality to conventional vascular imaging techniques. TaiPPG operates indirectly, without producing angiography, making it challenging to predict the specific location and scale of abnormalities. Consequently, it should not be solely relied upon for clinical diagnosis. However, like PPG, TaiPPG excels in monitoring early disease stages before symptoms are manifested, a realm often inaccessible to conventional vascular imaging due to resource and expertise requirements[34-36]. TaiPPG offers ubiquitous and frequent monitoring[37], facilitating early screening and risk detection[36,38,39]. Moreover, the TaiPPG-based prediction mechanism developed here expands the capabilities of existing PPG and mobile health techniques. While these methods primarily focus on global hemodynamic indicators, angiosome-resolved monitoring introduces additional data dimensions, enabling the extraction of parameters related to symmetry, evenness, phase differences, and relative intensities for the prediction of specific vascular conditions[40,41].

Nonetheless, our study does present certain limitations, which offer opportunities for future research endeavors. Firstly, the underlying mechanism of our observations remains speculative. While we have proposed a hypothesis in our discussion, additional validation with robust evidence is imperative. One promising avenue is ultrasound imaging, which can assess blood flow in the FA and even detect reverse flow in OA branches, shedding further light on our assumptions[42,43]. Additionally, studying the relationship between retinal characteristics and facial skin could provide valuable insights since OA perfusion may also affect retinal perfusion[44]. Thus, a future study focusing on the retina or the entire fundus would be highly valuable. Lastly, the TaiPPG system used in our study involved an advanced camera with a high bit depth, resulting in relatively higher costs and computational demands compared to current Internet of the Internet of medical things systems. To ensure wider accessibility, it will be crucial to optimize processing steps for deployment in low-resource settings.

In conclusion, we demonstrate that a facial imaging approach can quantitatively assess CT, offering significant clinical and public health implications for ocular health. These findings lay the scientific groundwork for mobile and digital health initiatives, allowing for the tracking of these traits in the general population without the need for extensive medical imaging.

## REFERENCES

1. Nickla, D.L. & Wallman, J. The multifunctional choroid. *Progress in retinal and eye research* **29**, 144-168 (2010).
2. Parver, L.M., Auker, C., & Carpenter, D.O. Choroidal blood flow as a heat dissipating mechanism in the macula. *American journal of ophthalmology* **89**, 641-646 (1980)
3. Grunwald, J.E., Metelitsina, T.I., DuPont, J.C., Ying, G.-S. & Maguire, M.G. Reduced foveolar choroidal blood flow in eyes with increasing AMD severity. *Investigative ophthalmology & visual science* **46**, 1033-1038 (2005).
4. Langham, M.E., Grebe, R., Hopkins, S., Marcus, S. & Sebag, M. Choroidal blood flow in diabetic retinopathy. *Experimental eye research* **52**, 167-173 (1991).
5. Ahn, S.M., *et al.* Retinal vascular flow and choroidal thickness in eyes with early age-related macular degeneration with reticular pseudodrusen. *BMC ophthalmology* **18**, 1-10 (2018).
6. Ahn, S.J., Ryu, S.J., Joung, J.Y. & Lee, B.R. Choroidal thinning associated with hydroxychloroquine retinopathy. *American Journal of Ophthalmology* **183**, 56-64 (2017).
7. Pang, C.E., Sarraf, D. & Freund, K.B. Extreme choroidal thinning in high myopia. *Retina* **35**, 407-415 (2015).
8. Garg, A., *et al.* Reticular pseudodrusen in early age-related macular degeneration are associated with choroidal thinning. *Investigative ophthalmology & visual science* **54**, 7075-7081 (2013).
9. Mullins, R.F., *et al.* The membrane attack complex in aging human choriocapillaris: relationship to macular degeneration and choroidal thinning. *The American journal of pathology* **184**, 3142-3153 (2014).
10. Read, S.A., Fuss, J.A., Vincent, S.J., Collins, M.J. & Alonso-Caneiro, D. Choroidal changes in human myopia: insights from optical coherence tomography imaging. *Clinical and Experimental Optometry* **102**, 270-285 (2019).
11. Kim, J.T., Lee, D.H., Joe, S.G., Kim, J.-G. & Yoon, Y.H. Changes in choroid thickness in relation to the severity of retinopathy and macular edema in type 2 diabetic patients. *Investigative ophthalmology & visual science* **54**, 3378-3384 (2013).
12. Zhou, H., *et al.* Age-related changes in choroidal thickness and the volume of vessels and stroma using swept-source OCT and fully automated algorithms. *Ophthalmology Retina* **4**, 204-215 (2020).
13. Laviers, H. & Zambarakji, H. Enhanced depth imaging-OCT of the choroid: a review of the current literature. *Graefe's Archive for Clinical and Experimental Ophthalmology* **252**, 1871-1883 (2014).
14. Taylor, G.I. & Palmer, J.H. The vascular territories (angiosomes) of the body: experimental study and clinical applications. *British journal of plastic surgery* **40**, 113-141 (1987).
15. Taylor, G.I., Corlett, R.J. & Ashton, M.W. The functional angiosome: clinical implications of the anatomical concept. *Plastic and reconstructive surgery* **140**, 721-733 (2017).
16. Houseman, N.D., Taylor, G.I. & Pan, W.-R. The angiosomes of the head and neck: anatomic study and clinical applications. *Plastic and reconstructive surgery* **105**, 2287-2313 (2000).
17. Taylor, G.I. The angiosomes of the body and their supply to perforator flaps. *Clinics in plastic surgery* **30**, 331-342 (2003).
18. Kikuchi, S., *et al.* Laser speckle flowgraphy can also be used to show dynamic changes in the blood flow of the skin of the foot after surgical revascularization. *Vascular* **27**, 242-251 (2019).
19. Kawarada, O., *et al.* Effect of single tibial artery revascularization on microcirculation in the setting of critical limb ischemia. *Circulation: Cardiovascular Interventions* **7**, 684-691 (2014).
20. Metelitsina, T.I., *et al.* Foveolar choroidal circulation and choroidal neovascularization in age-related macular degeneration. *Investigative ophthalmology & visual science* **49**, 358-363 (2008).
21. An, L., Subhush, H.M., Wilson, D.J. & Wang, R.K. High-resolution wide-field imaging of retinal and choroidal blood perfusion with optical microangiography. *Journal of biomedical optics* **15**, 026011-026011 (2010).
22. Cheng, Y., Chu, Z. & Wang, R.K. Robust three-dimensional registration on optical coherence tomography angiography for speckle reduction and visualization. *Quantitative Imaging in Medicine and Surgery* **11**, 879 (2021).
23. Patel, J., Shetty, A., Vishwakarma, A. & Shelke, V. Automated Face Detection And Swapping In Video.
24. Zhao, J., *et al.* A new face feature point matrix based on geometric features and illumination models for facial attraction analysis. *Discrete and Continuous Dynamical Systems-S* **12**, 1065-1072 (2019).
25. Chang, T.-R. & Tsai, M.-Y. Classifying conditions of speckle and wrinkle on the human face: A deep learning approach. *Electronics* **11**, 3623 (2022).
26. Shiohara, K. & Yamasaki, T. Detecting deepfakes with self-blended images. 18720-18729.
27. He, Q., Sun, Z., Li, Y., Wang, W. & Wang, R.K. Spatiotemporal monitoring of changes in oxy/deoxy-hemoglobin concentration and blood pulsation on human skin using smartphone-enabled remote multispectral photoplethysmography. *Biomedical Optics Express* **12**, 2919-2937 (2021).
28. Zhou, H., *et al.* Attenuation correction assisted automatic segmentation for assessing choroidal thickness and vasculature with swept-source OCT. *Biomedical optics express* **9**, 6067-6080 (2018).
29. Zhang, Y., *et al.* Influence of Carotid Endarterectomy on Choroidal Perfusion: The INFLATE Study. *Ophthalmology Retina* (2023).
30. van Laar, P.J., van der Grond, J., Bremmer, J.P., Klijn, C.J.M. & Hendrikse, J. Assessment of the contribution of the external carotid artery to brain perfusion in patients with internal carotid artery occlusion. *Stroke* **39**, 3003-3008 (2008).
31. Fearn, S.J., Picton, A.J., Mortimer, A.J., Parry, A.D. & McCollum, C.N. The contribution of the external carotid artery to cerebral perfusion in carotid disease. *Journal of vascular surgery* **31**, 989-993 (2000).
32. Altinbas, N.K., *et al.* Effect of carotid artery stenting on ophthalmic artery flow patterns. *Journal of Ultrasound in Medicine* **33**, 629-638 (2014).

33. Geibprasert, S., Pongpech, S., Armstrong, D. & Krings, T. Dangerous extracranial–intracranial anastomoses and supply to the cranial nerves: vessels the neurointerventionalist needs to know. *American journal of neuroradiology* **30**, 1459-1468 (2009).
34. Zahedi, E., Jaafar, R., Ali, M.A.M., Mohamed, A.L. & Maskon, O. Finger photoplethysmogram pulse amplitude changes induced by flow-mediated dilation. *Physiological measurement* **29**, 625 (2008).
35. Selvaraj, N., Jaryal, A.K., Santhosh, J., Anand, S. & Deepak, K.K. Monitoring of reactive hyperemia using photoplethysmographic pulse amplitude and transit time. *Journal of clinical monitoring and computing* **23**, 315-322 (2009).
36. Gandhi, P.G. & Rao, G.H.R. The spectral analysis of photoplethysmography to evaluate an independent cardiovascular risk factor. *International journal of general medicine*, 539-547 (2014).
37. Lao, C.K.*, et al.* Portable heart rate detector based on photoplethysmography with android programmable devices for ubiquitous health monitoring system. *International Journal of Advances in Telecommunications, Electrotechnics, Signals and Systems* **2**, 18-26 (2012).
38. Spigulis, J., Kukulis, I., Fridenberga, E. & Venckus, G. Potential of advanced photoplethysmography sensing for noninvasive vascular diagnostics and early screening. Vol. 4625 38-43 (SPIE).
39. Liang, Y., Chen, Z., Ward, R. & Elgendi, M. Hypertension assessment using photoplethysmography: a risk stratification approach. *Journal of clinical medicine* **8**, 12 (2018).
40. Kamshilin, A.A., Miridonov, S., Teplov, V., Saarenheimo, R. & Nippolainen, E. Photoplethysmographic imaging of high spatial resolution. *Biomedical optics express* **2**, 996-1006 (2011).
41. Zaunseder, S., Trumpp, A., Ernst, H., Förster, M. & Malberg, H. Spatio-temporal analysis of blood perfusion by imaging photoplethysmography. Vol. 10501 178-191 (SPIE).
42. Zhao, Z.*, et al.* Color Doppler flow imaging of the facial artery and vein. *Plastic and reconstructive surgery* **106**, 1249-1253 (2000).
43. Ward, J.B., Hedges Iii, T.R. & Heggerick, P.A. Reversible abnormalities in the ophthalmic arteries detected by color Doppler imaging. *Ophthalmology* **102**, 1606-1610 (1995).
44. Hwang, G.J.*, et al.* Reversal of ischemic retinopathy following balloon angioplasty of a stenotic ophthalmic artery. *Journal of Neuro-Ophthalmology* **30**, 228-230 (2010).